\documentclass[aps,letterpaper,twocolumn,nofootinbib]{revtex4}

\usepackage{graphicx} 	    
\usepackage{amsmath} 	   
\usepackage{amssymb}       
\usepackage{amsfonts}

\begin{document}

\title{Fractional Degrees of Freedom at Infinite Coupling in large Nf QED in 2+1 dimensions}

\author{Paul Romatschke}
\affiliation{Department of Physics, University of Colorado Boulder, CO 80309}

\begin{abstract}
  I consider quantum electrodynamics with many electrons in 2+1 space-time dimensions at finite temperature. The relevant dimensionless interaction parameter for this theory is the fine structure constant divided by the temperature. The theory is solvable at any value of the coupling, in particular for very weak (high temperature) and infinitely strong coupling (corresponding to the zero temperature limit). Concentrating on the photon, each of its physical degrees of freedom at infinite coupling only contributes half of the free-theory value to the entropy. These fractional degrees of freedom are reminiscent of what has been observed in other strongly coupled systems (such as ${\cal N}=4$ SYM), and bear similarity to the fractional Quantum Hall effect, potentially suggesting connections between these phenomena. The results found for QED3 are fully consistent with the expectations from particle-vortex duality.
\end{abstract}

\maketitle

\section{Introduction}

Quantum electrodynamics (QED) is one of the most successful theories in physics. Describing the interaction of matter and light, it is extremely well tested, achieving agreement with experiment on the parts-per-billion level for instance for the anomalous magnetic moment of the electron \cite{gm2}. Theory solutions to QED in three space and one time dimensions (QED4) are typically obtained by calculating a perturbative series in the fine structure constant $\alpha$. For comparison to experimental results such as those in Ref.~\cite{gm2} this strategy is perfectly adequate since the fine structure constant in natural units is $\alpha\simeq \frac{1}{137}$, so that power corrections of $\alpha$ are small.

However, it has been argued that the perturbative series for QED4 is divergent \cite{dyson}, implying that the theory becomes ambiguous at very high values of $\alpha$. Therefore, unfortunately, QED4 does not seem to be a suitable candidate if wanting to study a theory that is well-defined also in the limit of very strong (infinite) coupling.

Fortunately, QED does become well-defined if one is willing to reduce the number of space dimensions to two. In this 2+1 dimensional ``flatland'' case, QED3 is still formally defined by the same Lagrangian as its successful cousin QED4, but is apparently well behaved for any interaction strength. In particular, when considering massless QED3 in the limit of many electrons, QED3 becomes an interesting solvable theory to study, as has been the case for many years, cf. Refs.~\cite{Pisarski:1984dj,Appelquist:1988sr,Franz:2002qy,Giombi:2016fct}. One of the main differences with respect to QED4 is that the fine-structure constant $\alpha$ becomes dimensionful. Absent any mass-scale, the only dimensionless coupling for QED3 is thus the ratio of $\alpha$ and temperature, $\lambda\equiv\frac{\alpha}{T}$. It is possible to use units where $\alpha=1$, such that the weak coupling regime $\lambda \ll 1$ corresponds to the high temperature limit, and conversely the strong coupling regime $\lambda \gg 1$ corresponds to the zero temperature limit. This is a common feature of pure conformal field theories in 2+1 dimensions, see e.g. the discussion in Ref.~\cite{DeWolfe:2019etx}.

QED3 in the large number of electrons and large coupling limit $\lambda\rightarrow \infty$ is a strongly-coupled gauge theory with many components. The conjectured duality of strongly coupled gauge theories and gravity \cite{Maldacena:1997re} opened up the possibility of studying certain gauge theories in the large number of component and large coupling limit by mapping them to classical gravity. While QED3 does not have a known gravity dual, it is nevertheless interesting if some of the features found for holographic theories could be understood or recovered by performing calculations purely on the field theory side at strong coupling. This provides further motivation to study QED3 with many electrons.

\section{Calculation}

The Euclidean action for massless QED in 2+1 dimensions with $N_f$ electrons is given by \cite{Laine:2016hma}
\begin{eqnarray}
  \label{eq:segf}
  S_E=\int d^Dx \left[\frac{1}{4}F_{\mu\nu}F_{\mu\nu}+\bar\psi_a\gamma_\mu\left(\partial_\mu-i e A_\mu \right)\psi_a\right.\nonumber\\
    \left.+\frac{1}{2\xi}\left(\partial_\mu A_\mu\right)^2+\partial_\mu \bar c\partial_\mu c \right]\,,
  \end{eqnarray}
where $F_{\mu\nu}=\partial_\mu A_\nu-\partial_\nu A_\mu$ is the photon field strength tensor, $A_\mu$ is the U(1) gauge field, $\psi_a$ with $a=1,2,3,\ldots,N_f$ are massless Dirac fields and $\bar c,c$ are the Faddeev-Popov ghosts. Here $\xi$ is the gauge-fixing parameter in the class of covariant gauges considered, $D=3-2\epsilon$ is the dimension of the field theory with $\epsilon\geq 0$ in dimensional regularization, $\gamma_\mu$ are the Euclidean version of the Dirac $\gamma$ matrices and the relation of the coupling $e$ to the fine structure constant for $N_f$ electrons is taken to be $\alpha\equiv \frac{e^2 N_f}{4\pi}$. Note $S_E$ is invariant under BRST transformations.

In the large $N_f$ limit, the only contribution to the photon polarization tensor arises from the fermion loop, 
\begin{eqnarray}
   \label{eq:matterpi}
   \Pi_{\mu\nu}(P)=-16 \pi \alpha T \sum_n\int\frac{d^{D-1}k}{(2\pi)^{D-1}}\left[\delta_{\mu\nu}\left(K^2-P\cdot K\right)\right.\nonumber\\
     \left.-2 K_\mu K_\nu+K_\mu P_\nu+K_\nu P_\mu\right]\left[K^2(K-P)^2\right]^{-1}\,,\nonumber
\end{eqnarray}
where $K\equiv(\tilde \omega_n,{\bf k})$ and $\tilde \omega_n=2\pi T (n+\frac{1}{2})$ with $n\in \mathbb{Z}$ are the fermionic Matsubara modes. In the zero temperature limit, evaluation of $\Pi_{\mu\nu}$ can be found in many textbooks on quantum field theory, with the only change being $D=4-2\epsilon \rightarrow 3-2\epsilon$ \cite{Pisarski:1984dj}:
\begin{equation}
  \label{eq:pivacd3}
  \Pi_{\mu\nu}^{T=0}(P)=\frac{\alpha \pi}{2} \left(\delta_{\mu\nu}-\frac{P_\mu P_\nu}{P^2}\right) \sqrt{P^2}\,.
\end{equation}
At finite temperature, the presence of a preferred rest-frame introduces an additional vector $n_\mu\equiv(1,0,0)$, adding new tensor structures that $\Pi_{\mu\nu}$ can be decomposed in. Defining $\tilde n_\mu\equiv n_\mu\left(\delta_{\mu\nu}-\frac{P_\mu P_\nu}{P^2}\right)$ one finds that
\begin{equation}
  \Pi_{\mu\nu}(P)=\Pi_A \left(\delta_{\mu\nu}-\frac{P_\mu P_\nu}{P^2}-\frac{\tilde n_\mu \tilde n_\nu}{\tilde n^2}\right)+ \Pi_B \frac{\tilde n_\mu \tilde n_\nu}{\tilde n^2}\,,
  \end{equation}
where for high temperature $\Pi_{A,B}$ may be evaluated analytically in the Hard-Thermal-Loop approximation \cite{Braaten:1990az} or numerically.

The partition function for QED3 may be evaluated in the path-integral approach as
\begin{equation}
  Z=\int{\cal D}A {\cal D}\bar\psi {\cal D}\psi {\cal D}\bar{c} {\cal D}c e^{-S_E}\,.
\end{equation}
The fermions are unmodified, contributing
\begin{eqnarray}
  Z_{\rm fermions}&=&\prod_{K}{\rm det}\left[i \gamma_\mu K_\mu\right]^{N_f}=\prod_{K}{\rm det}\left[K^2 {\bf 1}_{4\times 4}\right]^{N_f/2}\,,\nonumber\\
  &=&e^{2 N_f V \sum_{\tilde w}\int\frac{d^{D-1}k}{(2\pi)^{D-1}} \ln K^2}\,,
\end{eqnarray}
with $V$ the ``volume'' of 2-dimensional space. Similarly, the ghosts, being Grassmann fields obeying periodic boundary conditions give
\begin{equation}
  \label{eq:ghostZ}
  Z_{\rm ghosts}=e^{V \sum_{w}\int\frac{d^{D-1}k}{(2\pi)^{D-1}} \ln K^2}\,.
\end{equation}
The photon, being dressed by the polarization tensor $\Pi_{\mu\nu}$, gives rise to
\begin{equation}
  \label{eq:photonZ}
  Z_{\rm photon}=e^{-\frac{V}{2}\sum_w\int\frac{d^{D-1}k}{(2\pi)^{D-1}} \ln \left[\left(K^2+\Pi_A\right)\left(K^2+\Pi_B\right)\left(\frac{K^2}{\xi}\right)\right]}\,,
\end{equation}
with $\xi$ again the gauge-fixing parameter that appeared in $S_E$.

The partition function defines the free energy density
\begin{equation}
  f\equiv-\frac{T}{V}\ln Z\,,
\end{equation}
at finite temperature. Basic thermodynamic relations then allow calculation of the entropy density as
\begin{equation}
  s\equiv -\frac{\partial f}{\partial T}\,.
\end{equation}

For instance, he free energy density for the fermions is given by
\begin{eqnarray}
    f_{\rm fermion}&=&-\frac{T}{V}\ln Z_{\rm fermion}\nonumber\\
    &=&-4 N_f T \int\frac{d^{D-1}k}{(2\pi)^{D-1}} \ln \left(1+e^{-k/T}\right)\,,
\end{eqnarray}
where I have performed the fermionic Matsubara sum and used the fact that divergent integrals without any inherent scale vanish in dimensional regularization, e.g. $\int\frac{d^{D}k}{(2 \pi)^D}\ln K^2=0$. The remaining integral is finite so that the limit $D\rightarrow 3$ may be taken. This leads to $f_{\rm fermion}=-\frac{3 N_f \zeta(3) T^3}{2\pi}$ or an entropy density for the $4N_f$ fermionic degrees of freedom of
\begin{equation}
  s_{\rm fermion}=-\frac{\partial f_{\rm fermion}}{\partial T}=\frac{9 N_f \zeta(3) T^2}{2\pi}\,.
\end{equation}
Similarly, the entropy density for the ghosts becomes
\begin{equation}
  s_{\rm gh}= \frac{\partial}{\partial T} \left[2 T \int\frac{d^{2}k}{(2\pi)^{2}} \ln \left(1-e^{-k/T}\right)\right]=-\frac{3\zeta(3)T^2}{\pi}\,.
\end{equation}
Finally, the free-energy contributions for the photon have the form
\begin{equation}
  f=\frac{T}{2}\sum_\omega \int \frac{d^{D-1}k}{(2\pi)^{D-1}}\ln \left[\omega_n^2+k^2+\Pi(\omega_n,k)\right]\,.
\end{equation}
For completeness, note that when $\Pi$ can be neglected, this gives rise to the entropy density of a single bosonic degree of freedom in 2+1 dimensions
\begin{equation}
  \label{eq:free}
  s_{\rm free}=-\frac{\partial}{\partial T} \left[ T \int\frac{d^{2}k}{(2\pi)^{2}} \ln \left(1-e^{-k/T}\right)\right]=\frac{3\zeta(3)T^2}{2\pi}\,.
\end{equation}

\section{Fractionalization of Photon Degrees of Freedom in QED3}

Inspecting Eq.~(\ref{eq:photonZ}), the photon field at finite temperature has three degrees of freedom. Let's call them ``A'', ``B'' and ``C''. Not all of these degrees of freedom are physical. In modern quantum field theory formalism, this comes about by the contribution from the Faddeev-Popov ghost (\ref{eq:ghostZ}), which is negative, and amounts to subtracting two bosonic degrees of freedom.

One degree of freedom from the ghosts exactly cancels one degree of freedom from the photon field (``C''), whereas the other ghost contribution only partially cancels one of the photon contributions because the photon acquires an in-medium mass and width. In QED3 in the limit of many electrons $N_f\gg 1$, this can be made exact by calculating the entropy density (entropy per ``volume'') of photons and ghosts,
\begin{eqnarray}
  \label{eq:master}
  &s_A+s_B+s_{\rm gh}=&\\
  &-\frac{\partial}{\partial T} \frac{T}{2}\sum_{n}\int\frac{d^2{\bf k}}{(2 \pi)^2}\ln\left[\frac{\left(\omega_n^2+{\bf k}^2+\Pi_A\right)\left(\omega_n^2+{\bf k}^2+\Pi_B\right)}{\omega_n^2+{\bf k}^2}\right]\,,&\nonumber
\end{eqnarray}
cf. Refs.~\cite{Moore:2002md,Ipp:2003zr}. In (\ref{eq:master}), $\Pi_A(\omega_n,{\bf k})$ and $\Pi_B(\omega_n,{\bf k})$ are the in-medium photon polarizations of photon degree of freedom ``A'' and ``B'', respectively, and the contribution from the ghost degree of freedom can be identified as residing in the denominator inside the logarithm. The divergent integral in (\ref{eq:master}) is to be understood in the sense of dimensional regularization, as is standard in quantum field theory.

Let us study the contribution of a single photon degree of freedom to the entropy\,,
\begin{equation}
  s=-\frac{\partial}{\partial T}\left[\frac{T}{2}\sum_{n}\int\frac{d^2{\bf k}}{(2 \pi)^2}\ln\left(\omega_n^2+{\bf k}^2+\Pi(\omega_n,{\bf k})\right)\right]\,.
  \end{equation}
At weak coupling (high temperature) $\frac{\alpha}{T}\rightarrow 0$, the photon polarizations become small, $\Pi_{A,B}\rightarrow 0$ and the entropy density is given by (\ref{eq:free}).

Conversely, in the limit of strong coupling (zero temperature) $\frac{\alpha}{T}\rightarrow \infty$, the polarization tensor components are found to be given by Eq.~(\ref{eq:pivacd3})\footnote{Careful readers may object that in-medium pieces for $\Pi_{A,B}$ could be expected to contribute in the naive $N_f\rightarrow 0$ limit. However, for sufficiently large, but finite $N_f$, the naive $\Pi_{\rm medium}\sim \alpha T$ behavior is expected to get modified to $\Pi_{\rm medium}\sim \alpha T \left(\frac{T}{\alpha}\right)^{\#/N_f}$, where $\#$ is a number that requires non-perturbative evaluation, cf. Refs.~\cite{Pisarski:1984dj,Appelquist:1988sr}, effectively suppressing these in-medium corrections.}, such that each photon degree of freedom contributes 
\begin{eqnarray}
  s_{\rm strong}&=&-\frac{\partial}{\partial T}\left[\frac{T}{2}\sum_{n}\int\frac{d^2{\bf k}}{(2 \pi)^2}\ln\left(\frac{\alpha \pi}{2} \sqrt{\omega_n^2+{\bf k}^2}\right)\right]\,,\nonumber\\
%
%    \omega_n^2+{\bf k}^2\right.\right.\nonumber\\
%    &&\left.\left.\qquad\qquad\qquad\qquad\qquad +\right)\right]\,,\nonumber\\
   &=&\frac{3\zeta(3)T^2}{4\pi}=\frac{s_{\rm free}}{2}\,.
  \end{eqnarray}
Thus, each photon degree of freedom at infinite coupling contributes only a fraction ($\frac{1}{2}$) of the non-interacting value to the entropy. As a consequence, the total entropy density (including the leading ${\cal O}(N_f)$ contribution from the electrons) becomes
\begin{equation}
  s_{\rm QED3,strong}=\frac{9 N_f\zeta(3) T^2}{2 \pi}+0+{\cal O}(N_f^{-1})\,,
  \end{equation}
since $s_A+s_B+s_{\rm gh}=0$ for $\frac{\alpha}{T}\rightarrow \infty$. Because the two ``fractionalized'' photon degrees of freedom cancel against the remaining ghost contribution, the entire ${\cal O}(N_f^0)$ contribution to the entropy vanishes, and the photon effectively has disappeared. Put differently, as far as degrees of freedom contributing to the entropy are concerned, QED3 in the strong coupling limit becomes a theory of $N_f$ ``emergent'' non-interacting Dirac fermions. Note that this is strikingly similar to expectations from particle-vortex duality for $N_f=1$ ``cousins'' of QED considered in Refs.~\cite{Son:2015xqa,Karch:2016sxi} (see also Ref.~\cite{Karch:2016aux} for a proposal of particle-vortex duality for $N_f\gg 1$.)

Note that this curious ``fractionalization'' of the photon contribution to the entropy comes about even though the photon dispersion relation, calculated as the solution of $-\omega^2+{\bf k^2}+\frac{\alpha \pi}{2}\sqrt{-\omega^2+{\bf k}^2}=0$ on the principal Riemann sheet is linear,
\begin{equation}
  \label{eq:disprel}
  \omega=\pm |{\bf k}|\,,
\end{equation}
so the photon remains massless with zero width in the zero-temperature limit. 

\section{Fractionalization in Other QFTs and Discussion}

Similar fractionalizations in the number of degrees of freedom for quantum field theories at infinite coupling have been reported before. For instance, using the conjectured gravity dual, the total entropy in ${\cal N}=4$ super Yang-Mills theory (SYM, containing gauge fields, scalars and fermions) for infinite coupling and large N has been found to be exactly $\frac{3}{4}$ that of the free theory value \cite{Gubser:1998nz}. Since ${\cal N}=4$ SYM is only solvable at infinite coupling through its conjectured gravity dual, this $\frac{3}{4}$ fraction (and how it may come about as possible fractionalization of the individual gauge, scalar and fermionic degrees of freedom) is as yet unexplained.

For the bosonic O(N) model in 2+1 dimensions a ratio of $\frac{4}{5}$ was found \cite{Sachdev:1993pr,Drummond:1997cw,Romatschke:2019ybu}. The fractionalization of the N scalar degrees of freedom is realized through a finite in-medium mass $\Pi\propto {\rm const.}$ at infinite coupling that happens to be twice the logarithm of the golden ratio. Thus, the dispersion relation for the scalars in the strongly interacting O(N) model is modified, unlike the photon in QED3, cf. Eq.~(\ref{eq:disprel}).

In the supersymmetric O(N) Wess-Zumino model in 2+1 dimensions, the strong-weak ratio of the total entropy was found to be $\frac{31}{35}$ \cite{DeWolfe:2019etx}. The Wess-Zumino model contains equal amounts of scalar and fermionic degrees of freedom. The factor of $\frac{31}{35}$ comes about through the $\frac{4}{5}$ fractionalization of the scalar degrees of freedom (as in the bosonic O(N) model), while the fermions remain un-fractionalized (as in QED3 above). Since fermions in 2+1d only contribute $\frac{3}{4}$ per degree of freedom to the total entropy, fractionalization of only the scalars leads to
\begin{equation}
  \frac{\frac{4}{5}+\frac{3}{4}}{1+\frac{3}{4}}=\frac{31}{35}\,.
\end{equation}

Fractionalization of the degrees of freedom in the entropy in these examples clearly is more subtle than for the photon in QED3 outlined above, which may explain why it has not received more attention in the literature.

Taken together, it is hard to ignore the apparent similarity between the fractionalization of degrees of freedom in the entropy in these strongly coupled relativistic quantum field theories and the fractional quantum Hall effect \cite{fqh}. Further work is needed to illuminate this possible connection, and turn it into a predictive instrument.

\section*{Acknowledgments} 
This work was supported in part by the Department of Energy, DOE award No. DE-SC0017905. I would like to thank A.~Karch, M.~Mezei, R.~Nandkishore, R.~Pisarski, M.~S\"appi, D.~Tong and A.~Vuorinen for helpful discussions.

\bibliographystyle{unsrt}
\bibliography{fra}
\end{document}